\documentclass[twocolumn,superscriptaddress,amsmath,amssymb,showpacs]{revtex4-2}

\usepackage{graphicx}
\usepackage{color}
\usepackage[colorlinks=true,citecolor=blue, urlcolor = blue, linkcolor= blue]{hyperref}
\renewcommand{\phi}{\varphi}

\begin{document}


\title{High-throughput screening of strong electron-phonon couplings in ternary metal diborides}

\author{Renhai Wang}
 \affiliation{School of Physics and Optoelectronic Engineering, Guangdong University of Technology, Guangzhou 510006, China}
 
\author{Yang Sun}
 \affiliation{Department of Physics and Astronomy, Iowa State University, Ames, Iowa 50011, USA}
 \affiliation{Department of Applied Physics and Applied Mathematics, Columbia University, New York, NY, 10027, USA}
 
\author{Feng Zhang}
 \affiliation{Department of Physics and Astronomy, Iowa State University, Ames, Iowa 50011, USA}
 \affiliation{Ames Laboratory-USDOE, Iowa State University, Ames, Iowa 50011, USA}
 
 \author{Feng Zheng}
 \affiliation{Department of Physics, Xiamen University, Xiamen 361005, China}
 
 \author{Yimei Fang}
 \affiliation{Department of Physics, Xiamen University, Xiamen 361005, China}
 
 \author{Shunqing Wu}
 \affiliation{Department of Physics, Xiamen University, Xiamen 361005, China}
 
\author{Huafeng Dong}
 \affiliation{School of Physics and Optoelectronic Engineering, Guangdong University of Technology, Guangzhou 510006, China}

\author{Cai-Zhuang Wang}
 \affiliation{Department of Physics and Astronomy, Iowa State University, Ames, Iowa 50011, USA}
 \affiliation{Ames Laboratory-USDOE, Iowa State University, Ames, Iowa 50011, USA}
 
 \author{Vladimir Antropov}
 \affiliation{Department of Physics and Astronomy, Iowa State University, Ames, Iowa 50011, USA}
 \affiliation{Ames Laboratory-USDOE, Iowa State University, Ames, Iowa 50011, USA}
 
\author{Kai-Ming Ho}
 \affiliation{Department of Physics and Astronomy, Iowa State University, Ames, Iowa 50011, USA}





\date{July 15, 2022}

\begin{abstract}
We perform a high-throughput screening on phonon-mediated superconductivity in ternary metal diboride structure with alkali, alkaline earth, and transition metals. We find 17 ground states and 78 low-energy metastable phases. From fast calculations of zone-center electron-phonon coupling, 43 compounds are revealed to show electron-phonon coupling strength higher than that of MgB$_2$. An anti-correlation between energetic stability and electron-phonon coupling strength is identified. We suggest two phases, i.e., Li$_3$ZrB$_8$ and Ca$_3$YB$_8$, to be synthesized, which show reasonable energetic stability and superconducting critical temperature.
\end{abstract}

\maketitle


\section{Introduction}

Superconductivity has irreplaceable applications in many fields such as energy, medical care, transportation, and quantum computing. The search for new superconductors with a high critical temperature 
(\textit{T$_c$}) is always a major scientific task that can open the door to many future techniques. Since 2001, the discovery of remarkably high superconducting \textit{T$_c$} in MgB$_2$ \cite{1_nagamatsu2001superconductivity} has stimulated great interest in searching for phonon-mediated superconductors in the layered hexagonal metal diborides structures \cite{2_buzea2001review, 3_kortus2001superconductivity, 4_belashchenko2001coexistence, 5_an2001superconductivity, 6_liu2001beyond, 7_kong2001electron, 8_choi2002first, 9_mazin2003electronic}. Many attempts have been made by doping other elements to increase \textit{T$_c$} of MgB$_2$ \cite{10_cava2003substitutional, 11_karpinski2007single, 12_karpinski2008mgb}, which include Mg$_{1-x}$Li$_x$B$_2$ \cite{13_pallecchi2009investigation}, Mg$_{1-x}$Zr$_x$B$_2$ \cite{14_feng2002enhanced}, Mg$_{1-x}$Zn$_x$B$_2$ \cite{15_toulemonde2003high}, Mg$_{1-x}$Nb$_x$B$_2$ \cite{16_kalavathi2005superconductivity}, Mg$_{1-x}$Sc$_x$B$_2$ \cite{17_agrestini2004substitution}, Mg$_{1-x}$Ti$_x$B$_2$ \cite{18_lee2008doping}, Mg$_{1-x}$Na$_x$B$_2$ \cite{15_toulemonde2003high} and Mg$_{1-x}$Ca$_x$B$_2$ \cite{15_toulemonde2003high}. However, most experimental data show a decreasing trend of \textit{T$_c$} with an increasing amount of doping metal. Other metal diborides phases without Mg have also been explored, such as ZrB$_2$ \cite{19_gasparov2001electron}, NbB$_2$ \cite{20_leyarovska1979search} and TaB$_2$ \cite{21_rosner2001electronic}; however, they only showed vanishing \textit{T$_c$}. Recent experiments identify MoB$_2$ \cite{22_pei2021pressure} and WB$_2$ \cite{23_lim2021creating} showing high \textit{T$_c$} while high pressures of greater than 50 GPa are required. First-principle calculations also demonstrated a few possibilities of high \textit{T$_c$} in metal diborides. A well-known case is the CaB$_2$ \cite{24_choi2009prediction}, where the electron-phonon coupling (EPC) is much stronger than that in MgB$_2$. Unfortunately, it’s difficult to synthesize due to inferior thermodynamic stability. Doping with Cd and Ba in MgB$_2$ is predicted to show higher \textit{T$_c$} than MgB$_2$, but such doped structures are also difficult to be synthesized experimentally \cite{25_mackinnon2017phonon, 26_alarco2015phonon}. Therefore, the thermodynamic stability and superconducting properties should be considered simultaneously to find new experimentally feasible ternary superconductors in metal diborides.

While the dopants should be as diverse as possible, it is difficult to efficiently screen out promising superconductors among a large number of candidates with theoretical \textit{T$_c$} calculations. This is mainly because the detailed calculation of the electron-phonon coupling from density-functional perturbation theory \cite{27_baroni2001phonons} is complicated and time-consuming. We recently found that single-cell frozen-phonon calculations of the EPC strength of the zone-center phonons can be an efficient alternative to full density functional theory (DFT) evaluation of the Eliashberg function \cite{28_sun2022electron}. It well distinguishes strong EPC in the MgB$_2$ and high-pressure hydride systems \cite{28_sun2022electron}. Therefore it can be a fast descriptor of the full Brillouin zone EPC constant for the metal diboride family. In this work, we employ this method to perform a fast screening of strong EPC on ternary metal diboride phases. A large number of substituted phases are screened based on energetical stability and zone-center EPC strength. The detailed full Brioullion zone calculations of EPC and \textit{T$_c$} are performed for promising candidates.

\section{Results and Discussion}

\subsection{Binary metal diboride phases}

\begin{figure}[t]
\includegraphics[width=0.475\textwidth]{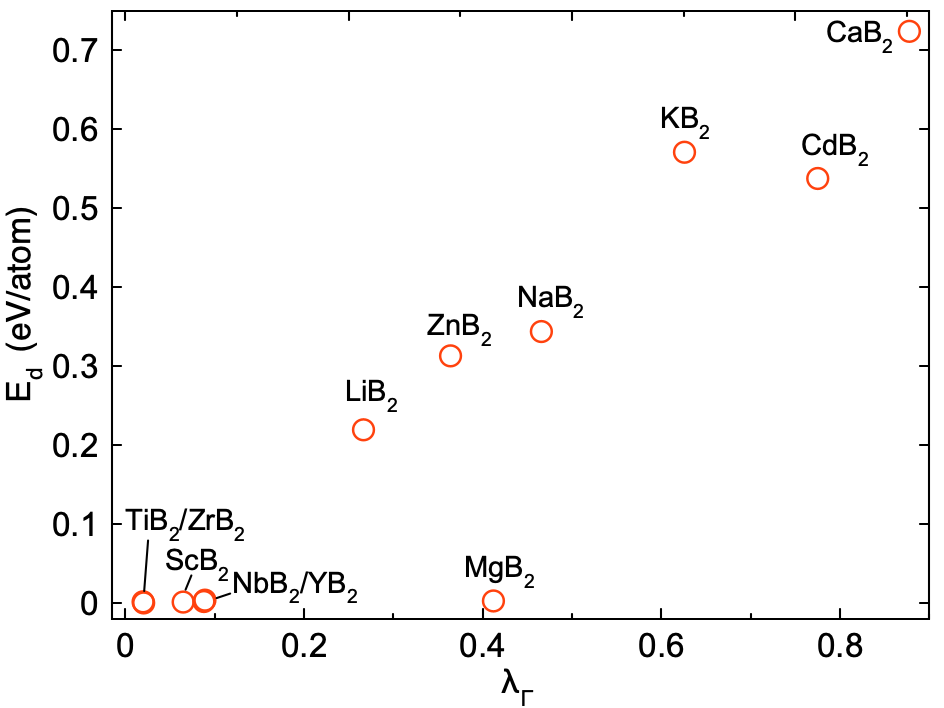}
\caption{\label{Fig1}
E$_d$-$\lambda_\Gamma$ diagram of 12 binary MB$_2$ structures. Lower E$_d$ indicates better stability. Higher $\lambda_\Gamma$ indicates stronger EPC.
}
\end{figure}

We first examine the effect of metal substitution in the binary metal diboride structure with 12 elements including Ca, Cd, K, Na, Zn, Li, Nb, Y, Sc, Zr, Ti and Mg. The energetic stability is described by the energy above the convex hull E$_d$ (see the method section for definition of E$_d$). The EPC strength is described with the zone-center EPC strength $\lambda_\Gamma$ (see the method section for definition of $\lambda_\Gamma$). As shown in Fig.~\ref{Fig1}, Zr, Ti, Sc, Y, Nb diborides show good energetic stability but no EPC. Ca, Cd, K diborides show strong EPC but poor energetic stability. An anti-correlation between E$_d$ and $\lambda_\Gamma$ can be seen in these binary phases. A stronger EPC (larger $\lambda_\Gamma$) leads to worse energetic stability (higher E$_d$). Interestingly, MgB$_2$ phase shows both a decent $\lambda_\Gamma$ and strong energetic stability, which is an outlier from the general anti-correlation of the two properties. CaB$_2$ system shows the best EPC but poor energetic stability, making it difficult to synthesize in experiments \cite{24_choi2009prediction}. Therefore the correlation in Fig.~\ref{Fig1} is in line with the previous findings that substitution of Mg in MgB$_2$ either leads to lower \textit{T$_c$} or reduced stability. This demonstrates that the $\lambda_\Gamma$ provides a good description of EPC strength for metal diboride systems, consistent with a previous comparative study between MgB$_2$ and AlB$_2$ \cite{28_sun2022electron}.
 
 \subsection{M$_{1-x}$N$_x$B$_2$ ternary convex hull}
 
\begin{figure}[t]
\includegraphics[width=0.3\textwidth]{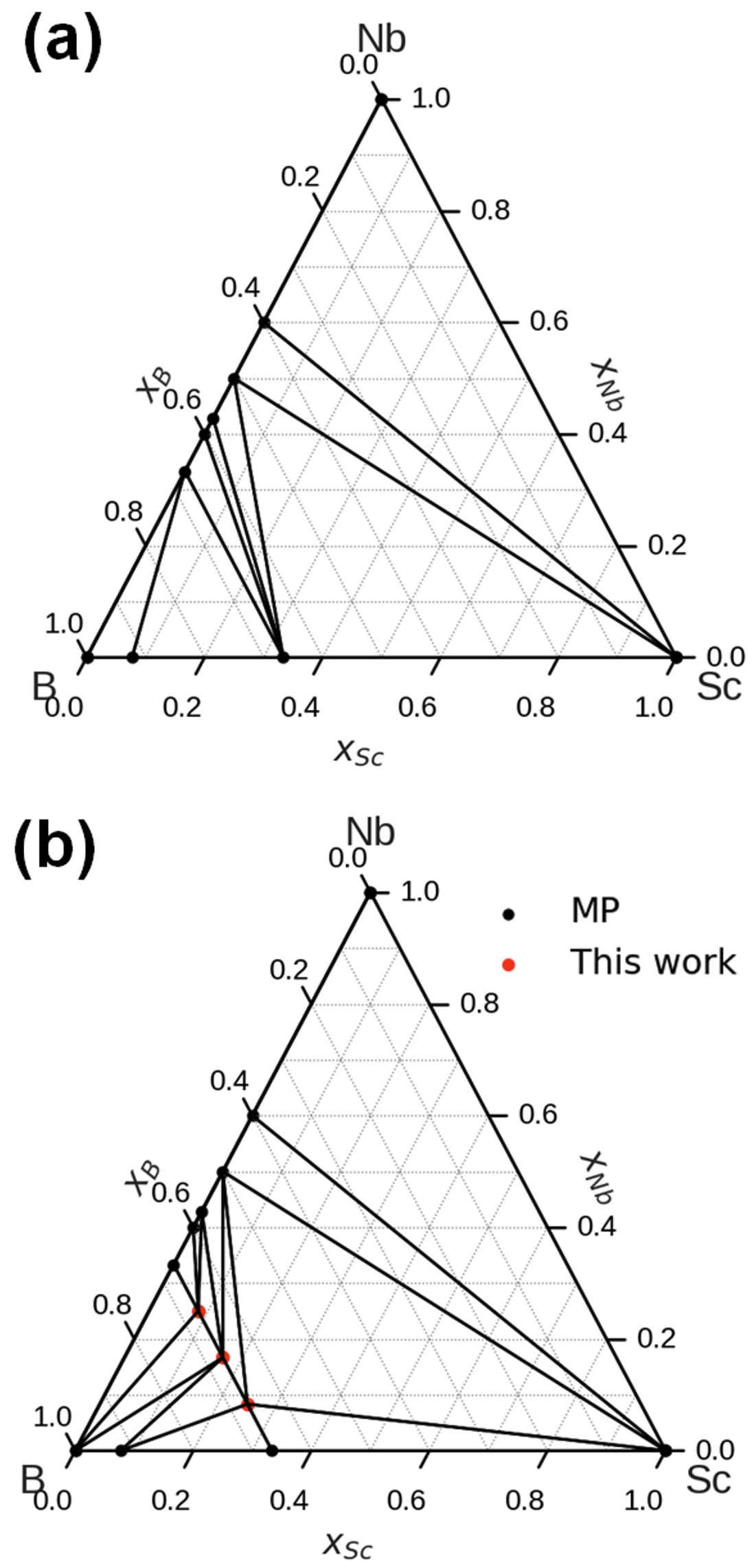}
\caption{\label{Fig2}
Convex hull of the Sc-Nb-B system. (a) Previously reported phases from MP database; (b) Convex hull including the ScNbB$_4$, ScNb$_3$B$_8$ and Sc$_3$NbB$_8$ phases. The black lines separate the compositional space to Gibbs triangles.
}
\end{figure}

 We calculate energetic stability and EPC strength in the ternary M$_{1-x}$N$_x$B$_2$ by mixing 12 metals on the metal sites in the supercell of the AlB$_2$ structure (see Method). The combination of M and N results in 66 ($C_{12}^2$) M-N-B ternary systems with 198 geometrially inequivalent structures containing three compositions, i.e., MNB$_4$, MN$_3$B$_8$ and M$_3$NB$_8$. The convex hull is constructed for each ternary phase to describe its energetic stability. Taking Nb-Sc-B as an example in Fig.~\ref{Fig2}, the compositional space (Gibbs triangle) is partitioned into multiple triangular pieces by the ground state structures, which form the corners of the convex hull for the corresponding ternary system. The known stable ground states are obtained from the Materials Project (MP) \cite{29_jain2013commentary} database. In Fig.~\ref{Fig2}(a), the known Na-Sc-B phases form a reference convex hull. If any new structure has formation energy below this convex hull surface, it is defined as a new ground state, and the convex hull surface is updated by including the new phase. In Fig.~\ref{Fig2}(b), three Nb-Sc-B structures (ScNbB$_4$, ScNb$_3$B$_8$, Sc$_3$NbB$_8$) are found to be the ground states phases, considerally modifying the original convex hull reference shown in Fig.~\ref{Fig2}(a).
 
In total, 17 ternary phases are identified as ground states, including ScNbB$_4$, ScNb$_3$B$_8$, Sc$_3$NbB$_8$, LiNb$_3$B$_8$, CaNbB$_4$, YNbB$_4$, YZrB$_4$, ZrScB$_4$, Zr$_3$ScB$_8$, ZrNbB$_4$, ScTiB$_4$, Mg$_3$ScB$_8$, MgNbB$_4$, MgNb$_3$B$_8$, Ti$_3$NbB$_8$, TiNbB$_4$ and TiNb$_3$B$_8$. Two of them, i.e., TiNbB$_4$ and ZrNbB$_4$, are previously reported in the MP database. Moreover, Mg$_{0.75}$Sc$_{0.25}$B$_2$ \cite{17_agrestini2004substitution} and three Zr$_{1-x}$Nb$_x$B$_2$ (x=0.25, 0.5, 0.75) were experimentally synthesized \cite{30_akarsu2021effects}. The convex hulls are updated by including the new ground states, shown in Supplementary Materials Fig. S2. Besides, we identify many low-energy metastable states that may be synthesizable by experiments, especially under non-equilibrium synthesis routes. For instance, recent experiments on LiNiB showed that the Li$_{0.75}$[NiB]$_2$ phase with E$_d$=0.21 eV/atom can be synthesized from high-temperature reactions \cite{31_bhaskar2021topochemical}. Metastable SnTi$_2$N$_4$ (E$_d$=0.2 eV/atom) \cite{32_bikowski2017design} and metastable ZnMoN$_2$ in a wurtzite-derived structure (E$_d$=0.16 eV/atom) \cite{33_arca2018redox} are all successfully synthesized in experiments. Here, by using a threshold of $E_d < 0.2$ eV/atom, we identify 78 metastable metal diboride phases, which may have experimental synthesizability. The detailed information on these metastable phases is shown in Supplementary Table S1.

\subsection{E$_d$-$\lambda_\Gamma$ correlation in ternary M$_{1-x}$N$_x$B$_2$ }

\begin{figure}[t]
\includegraphics[width=0.49\textwidth]{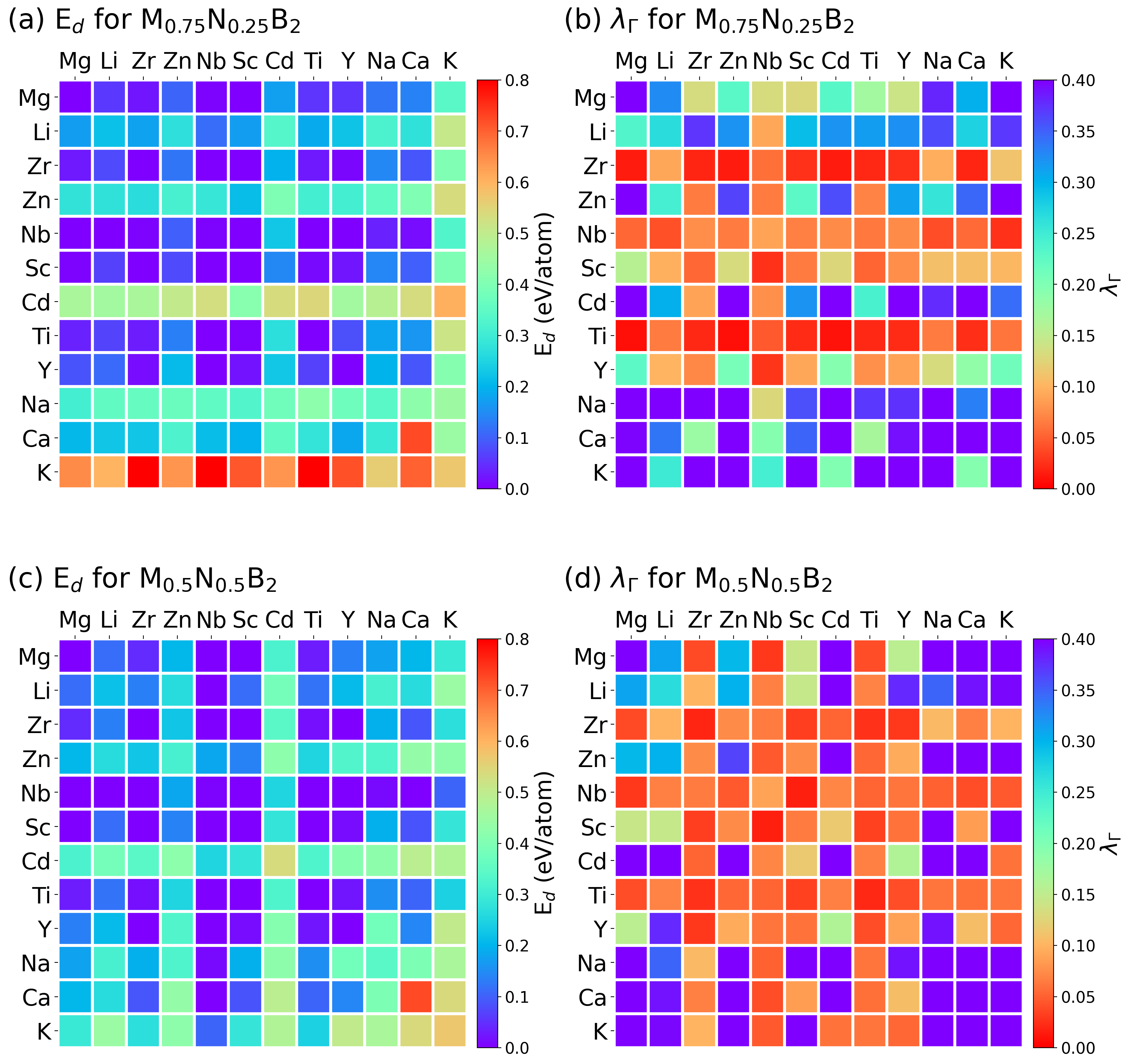}
\caption{\label{Fig3}
E$_d$ and $\lambda$ mapping of ternary metal diboride compounds. (a) and (b) are Ed and $\lambda$ for M$_{0.75}$N$_{0.25}$B$_2$; (c) and (d) are E$_d$ and $\lambda$ for M$_{0.5}$N$_{0.5}$B$_2$. X axis indicates N site and Y axis indicates M site.
}
\end{figure}

To describe EPC, $\lambda_\Gamma$ are computed for all ternary M$_{1-x}$N$_x$B$_2$. The maps of E$_d$ and $\lambda_\Gamma$ with all substituted structures are shown in Fig.~\ref{Fig3}. The color of each grid indicates the E$_d$ or $\lambda$ value of the corresponding phases. The more bluish coding indicates better energetic stability or stronger EPC. The data of the binary phases are listed in the diagonal grids as reference. One can see the compounds containing Mg, Zr, Nb, Sc, Ti, and Y elements show better thermodynamic stability while those containing Mg, Li, Cd, Na, Ca and K have better EPC strength. 

\begin{figure}[t]
\includegraphics[width=0.48\textwidth]{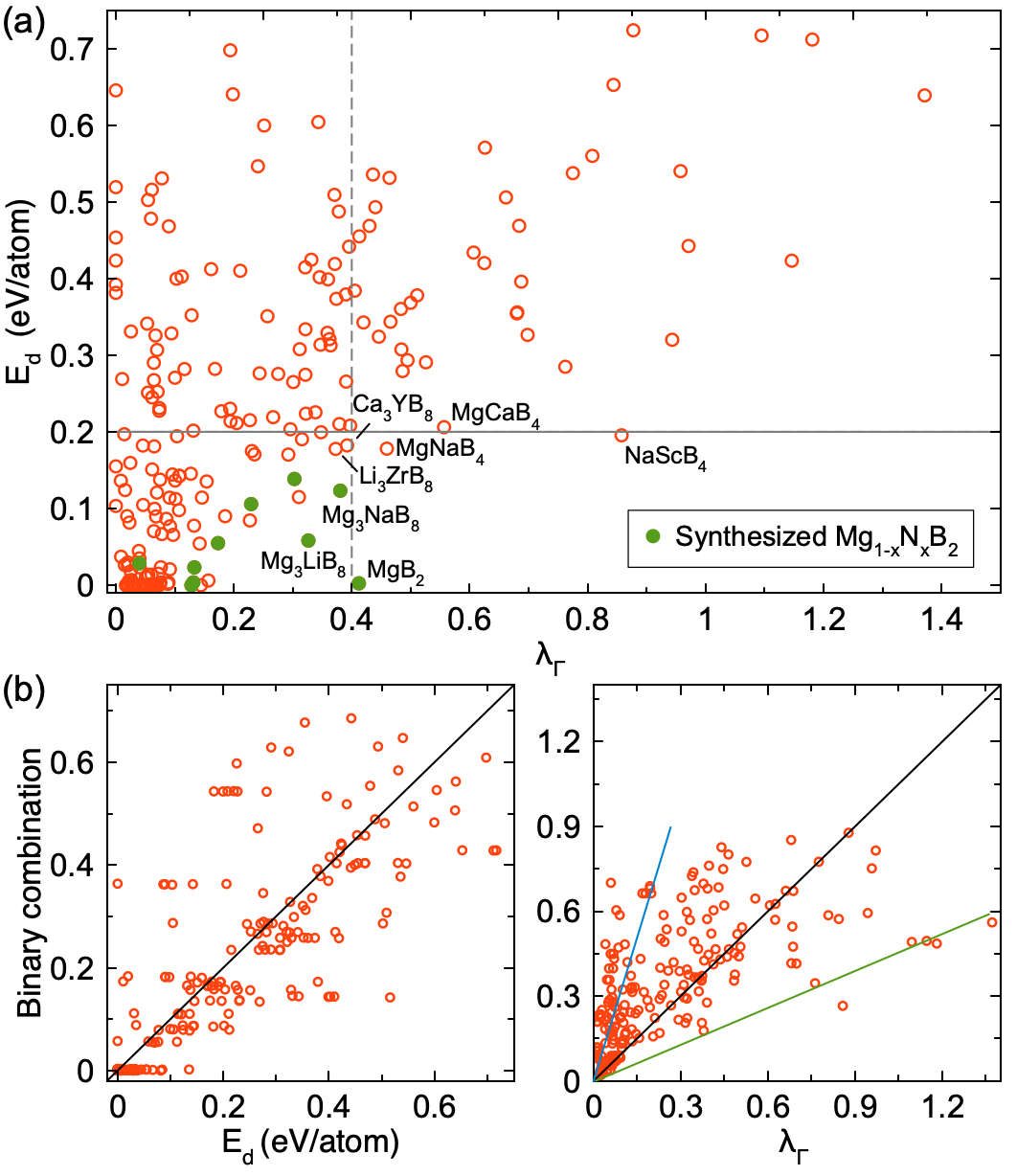}
\caption{\label{Fig4}
E$_d$ and $\lambda_\Gamma$ correlation in ternary metal diborides. (a) E$_d$-$\lambda_\Gamma$ diagram for ternary M$_{1-x}$N$_x$B$_2$ structures. Green solid symbol indicates phases previously synthesized by experiments \cite{1_nagamatsu2001superconductivity, 13_pallecchi2009investigation, 14_feng2002enhanced, 15_toulemonde2003high, 16_kalavathi2005superconductivity, 17_agrestini2004substitution, 18_lee2008doping}. The horizontal line indicates the range of synthesizable energetic stability. The vertical dashed line indicates $\lambda_\Gamma$ value similar to MgB$_2$. (b) The comparison between the E$_d$ (left pannel) and $\lambda_\Gamma$ (right panel) of ternary phases and the linear combinations of their binary counterparts. The x axis shows the value of tenary phases, i.e., E$_d$(M$_{1-x}$N$_x$B$_2$) or $\lambda_\Gamma$(M$_{1-x}$N$_x$B$_2$). The y axis shows the linear combination of the binary counterparts, i.e. (1-x)E$_d$(MB$_2$)+xE$_d$(NB$_2$) or (1-x)$\lambda_\Gamma$(MB$_2$)+x$\lambda_\Gamma$(NB$_2$). The black line indicates y=x correlation. Blue and green lines indicate the deterioration and enhancement of $\lambda_\Gamma$, respectively.
}
\end{figure}

The E$_d$-$\lambda_\Gamma$ correlation of all ternary phases are plotted in Fig.  \ref{Fig4}(a), which also shows a general trend of anti-correlation between stability and EPC strength, although the ternary data points appear to be more scattered than the binary ones shown in Fig. \ref{Fig1}. To understand the cation mixing effect in ternary phases, we recalculate E$_d$ and $\lambda_\Gamma$ values by linearly interpolating between binary phases (e.g., E$_d$(M$_{1-x}$N$_x$B$_2$)=(1-x)E$_d$(MB$_2$)+xE$_d$(NB$_2$)), and compare the interpolated values with the real values in Fig. \ref{Fig4}(b). E$_d$ mostly follow the y=x line, indicating insignificant mixing enthalpy for the ternary phases. However, $\lambda_\Gamma$ in Fig. \ref{Fig4}(b) strongly deviates from the y=x line. A large group of ternary phases show a deteriorated $\lambda_\Gamma$ compared to the linear combination of parent phases (blue line in Fig.  \ref{Fig4}(b)). Only a small group of ternary phases show an enhancement of $\lambda_\Gamma$ due to the mixing (green line). This provides a qualitative explanation to many previously failed attempts at increasing \textit{T$_c$} of MgB$_2$ by doping with other elements. Because the substitution has a much higher chance of deteriating, instead of enchancing, the EPC. Therefore, it is necessary to perform a high-throughput screening of many substitution possibilities. We also checked the correlation between the density of states at Fermi level N($\epsilon_F$) and E$_d$ or $\lambda_\Gamma$, and verified that N($\epsilon_F$) is not a dominant factor to fully describe the total energy and EPC.

\begin{figure*}[t]
\includegraphics[width=0.75\textwidth]{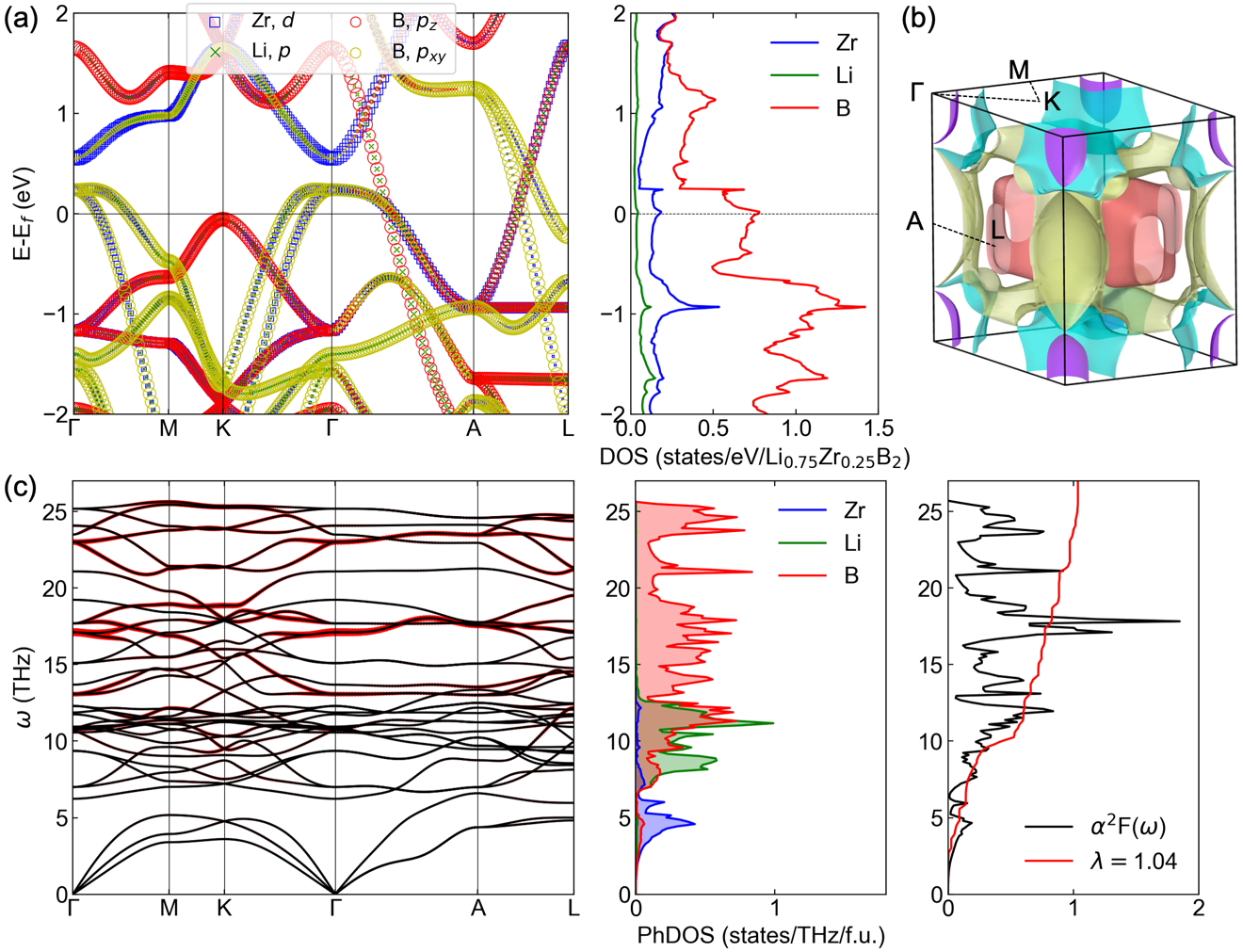}
\caption{\label{Fig5}
Electron structure and electron-phonon calculations for Li$_3$ZrB$_8$. (a) Electronic band structure and projected density of states. (b) Fermi surface. (c) Phonon dispersion, phonon density of state and Eliashberg spectrum. The red bands on the phonon dispersion indicate the strength of EPC.
}
\end{figure*}

A few ternary metal diborides containing Mg have been successfully synthesized previously \cite{1_nagamatsu2001superconductivity, 13_pallecchi2009investigation, 14_feng2002enhanced, 15_toulemonde2003high, 16_kalavathi2005superconductivity, 17_agrestini2004substitution, 18_lee2008doping}, as marked in Fig. \ref{Fig4}. Table \ref{Tab1} lists detailed information on the doped Mg$_{1-x}$N$_x$B$_2$ that have been reported experimentally or calculated theoretically \cite{1_nagamatsu2001superconductivity, 13_pallecchi2009investigation, 14_feng2002enhanced, 15_toulemonde2003high, 16_kalavathi2005superconductivity, 17_agrestini2004substitution, 18_lee2008doping, 25_mackinnon2017phonon, 34_singh2003theoretical, 35_neaton2001possibility}. One can see $\lambda_\Gamma$  shows a consistent trend with the reported \textit{T$_c$} values, indicating that the zone-center EPC calculation can reliably estimate the superconducting properties of this material family. It’s important to note that experiments show that 20\% of Zn, 10\% of Na and 10\% of Ca can be doped in the MgB$_2$ structure \cite{15_toulemonde2003high}, while their E$_d$  values are in the range of 0.1-0.2 eV/atom. Therefore, it provides a range of E$_d$ to identify metastable phases that may be accessible in experiments. In Fig. \ref{Fig4}, we use $E_d < 0.2$ eV/atom as the criteria to select structures with good stability. We use a threshold of $\lambda_\Gamma \sim 0.4$, (i.e.$\sim \lambda_\Gamma$ of MgB$_2$) to screen out phases with good EPC. The phases of interest that satisfy both criteria are located in the lower-right area of Fig. \ref{Fig4}. While most phases in or near this area are the doped MgB$_2$, it also identifies three non-MgB$_2$ phases for further study, namely NaScB$_4$, Li$_3$ZrB$_8$ and Ca$_3$YB$_8$. By checking the full Brillouin zone phonon spectrum, we find NaScB$_4$ shows strong imaginary phonons (see Supplementary Fig. S3) while Li$_3$ZrB$_8$ and Ca$_3$YB$_8$ are dynamically stable. By analyzing the zone-center phonon modes, we find the EPCs in both systems are contributed by Raman-active E$_{2g}$ modes, as shown in Fig. S4. These modes are two-dimensional on the boron layers, similar to the stretching modes in MgB$_2$, while the distributions on boron atoms are different (Fig. S4). It’s interesting to note that $\lambda_\Gamma$ of Li$_3$ZrB$_8$ is higher than those of LiB$_2$ and ZrB$_2$. Therefore, Li$_3$ZrB$_8$ can be an example of the enhanced EPC group due to the mixing (green line in Fig. \ref{Fig4} (b)).

\begin{table*}[t]
  \caption{Zone-center EPC strength $\lambda_\Gamma$, previously calculated \textit{T$_c$} or experimental \textit{T$_c$} and energy above the convex hull E$_d$ from present calculations for doped Mg$_{1-x}$N$_x$B$_2$ phases. The $x$ in the bracket shows the previously studied composition.}
  \label{Tab1}
  \begin{ruledtabular}
  \begin{tabular}{ccccc}
    Mg$_{1-x}$N$_x$B$_2$   & $\lambda_\Gamma$   & Calculated \textit{T$_c$} (K)   & Experimental \textit{T$_c$} (K)  & E$_d$ (eV/atom)     \\
    \hline
    MgB$_2$            & 0.41  & 42 \cite{26_alarco2015phonon} & 39 \cite{1_nagamatsu2001superconductivity} & 0.0         \\
    Mg$_{0.75}$Nb$_{0.25}$B$_2$  & 0.13  & - & 39.3 ($x$=0.05) \cite{16_kalavathi2005superconductivity} & 0.004         \\
    Mg$_{0.75}$Li$_{0.25}$B$_2$  & 0.33  & 31 ($x$=0.2) \cite{34_singh2003theoretical} & 38.3 ($x$=0.3) \cite{13_pallecchi2009investigation} & 0.058         \\
    Mg$_{0.75}$Na$_{0.25}$B$_2$  & 0.38  & 44-54 ($x$=0.2) \cite{35_neaton2001possibility} & 38 ($x$=0.1) \cite{15_toulemonde2003high} & 0.123   \\
    Mg$_{0.75}$Ca$_{0.25}$B$_2$  & 0.31  & 41-52 ($x$=0.2) \cite{35_neaton2001possibility} & 38 ($x$=0.1) \cite{15_toulemonde2003high} & 0.139   \\  
    Mg$_{0.75}$Zn$_{0.25}$B$_2$  & 0.23  & 33 ($x$=0.2) \cite{34_singh2003theoretical} & 38 ($x$=0.2) \cite{15_toulemonde2003high} & 0.106       \\
    Mg$_{0.75}$Zr$_{0.25}$B$_2$	& 0.13	& -	& 37.3 ($x$=0.2) \cite{14_feng2002enhanced} & 0.023   \\
    Mg$_{0.75}$Ti$_{0.25}$B$_2$	& 0.18	& 25.5 \cite{25_mackinnon2017phonon}    & 30 ($x$=0.2) \cite{18_lee2008doping}    & 0.055     \\
    Mg$_{0.75}$Sc$_{0.25}$B$_2$	& 0.13	& 11.4 \cite{25_mackinnon2017phonon}    & 8.2 \cite{17_agrestini2004substitution}   & 0.0   \\
    Mg$_{0.5}$Ti$_{0.5}$B$_2$	& 0.04	& 4.9 \cite{25_mackinnon2017phonon}     & 26 ($x$=0.4) \cite{18_lee2008doping}    &   0.029   \\
    Mg$_{0.5}$Sc$_{0.5}$B$_2$	& 0.15	& 8.8 \cite{25_mackinnon2017phonon}     & -	                                    &   0.003   \\
  \end{tabular}
  \end{ruledtabular}
\end{table*}

\subsection{Superconductivity in Li$_3$ZrB$_8$ and Ca$_3$YB$_8$}

Because of promising synthesizability and EPC in Li$_3$ZrB$_8$ and Ca$_3$YB$_8$ phases, we perform DFPT calculations to compute the full Brioullion zone EPC constant and calculate \textit{T$_c$} with McMillan equations (see Method). Fig. \ref{Fig5} shows the electronic structure and phonon spectrum for Li$_3$ZrB$_8$. The bands at the Fermi level are mainly from B’s $p$ electrons mixed with Zr’s $d$ electrons (Fig. \ref{Fig5}(a)). Compared to the electronic structure of MgB$_2$ (Supplementary Fig. S5), the flat bands from $\Gamma$ to A shows a cross at the Fermi level, which also results in a significant change in the Fermi surface in Fig. \ref{Fig5} (b).

We calculate \textit{T$_c$} for Li$_3$ZrB$_8$ and Ca$_3$YB$_8$ with McMillan equations. We also re-calculate \textit{T$_c$} for MgB$_2$ with the same method and same density of k- and q-grid (see Method section). This provides us a reference to estimate \textit{T$_c$} in the two ternary systems. For MgB$_2$, we obtain isotropic \textit{T$_c$}=19 K with the Allen-Dynes formula. This is consistent with the previous calculation (22 K in Ref. \cite{36_liu2001beyond}) but underestimates the \textit{T$_c$} compared to the experimental value of 39 K. The error is mainly due to the McMillan equation. One may improve it by employing a more sophisticated anisotropic Eliashberg theory \cite{24_choi2009prediction} and Migdal-Eliashberg equation \cite{37_margine2013anisotropic}. Nevertheless, using the same accuracy, we obtain \textit{T$_c$} as 38 K and 10 K for Li$_3$ZrB$_8$ and Ca$_3$YB$_8$, respectively. Therefore, Li$_3$ZrB$_8$ shows \textit{T$_c$} almost twice as larger than the one in MgB$_2$, while the \textit{T$_c$} of Ca$_3$YB$_8$ is half of MgB$_2$.

\section{Conclusion}

In summary, using first-principles high-throughput calculations, we search for ternary metal diborides with energetic stability and high EPC strength in 66 systems. 17 phases are identified to be stable ternary ground states and the ternary phase diagrams of these systems are updated accrordingly. 78 metastable phases with $E_d<0.2$ eV/atom are also identified. An anti-correlation between energetic stability and EPC strength is revealed in both binary and ternary metal diborides. Two systems, Li$_3$ZrB$_8$ and Ca$_3$YB$_8$, show both high synthesizability and strong EPC strength. The \textit{T$_c$} of Li$_3$ZrB$_8$ is predicted to be twice as large as that of MgB$_2$, calculated based on the  McMillan formulism with the same parameters. The experimental verification of our prediction is highly desirable. Our studies demonstrate zone-center phonon calculations as an encouraging method for massive screening of multi-component systems for conventional high-\textit{T$_c$} superconductors.

\section{Computational Methods}

The AlB$_2$-type primitive cell (space group: $P6/mmm$) was expanded by 1×1×2 or 2×2×1 to generate M$_{1-x}$N$_x$B$_2$ ternary metal diboride structures (M and N representing the metal elements). The ratio x includes 0.25 (M$_3$NB$_8$), 0.50 (MNB$_4$) and 0.75 (MN$_3$B$_8$). We consider unique substitutional sites, which generate one configuration for M$_3$NB$_8$ or MN$_3$B$_8$ and two for MNB$_4$ (see Supplementary Materials Fig. S1 for details). The ternary structures are optimized by \emph{ab initio} calculations, which were performed using the projector augmented wave (PAW) method \cite{38_blochl1994projector} within density functional theory as implemented in the VASP code \cite{39_kresse1996efficiency, 40_kresse1996efficient}. The exchange and correlation energy are treated without the spin-polarized generalized gradient approximation (GGA) and parameterized by the Perdew-Burke-Ernzerhof formula (PBE) \cite{41_perdew1996generalized}. A plane-wave basis was used with a kinetic energy cutoff of 520 eV, and the convergence criterion for the total energy was set to $10^{-5}$ eV. Monkhorst-Pack's sampling scheme \cite{42_monkhorst1976special} was adopted for Brillouin zone sampling with a k-point grid of $2{\pi} \times 0.033~{\rm{\AA}}^{-1}$. The lattice vectors (supercell shape and size) and atomic coordinates are fully relaxed until the force on each atom is less than 0.01 {eV/\AA}.

The formation energy E$_f$ of ternary M$_x$N$_y$B$_z$ is calculated by
\begin{equation}
   E_f = \frac{E({M_xN_yB_z}) - x E({M}) - y E({N}) - z E({B})}{x+y+z}
   \label{E_form}
\end{equation}
where $E(M_x N_y B_z )$ is the total energy of the M$_x$N$_y$B$_z$; $E(M)$, $E(N)$ and $E(B)$ are the total energy of M, N, and B ground-state bulk phases, respectively. To characterize the energetic stability of M$_x$N$_y$B$_z$, the formation energy differences with respect to the three reference phases forming the Gibbs triangle on the convex hull (denoted as E$_d$) are calculated. If E$_d$=0, it indicates the M$_x$N$_y$B$_z$ is a new ground state and the existing convex hull should be updated. The reference of convex hulls are obtained from the Material Project dataset \cite{29_jain2013commentary}. 

The high-throughput screening of strong EPC in these metal borides is based on fast frozen-phonon calculation of zone-center EPC strength \cite{28_sun2022electron}, defined by
\begin{equation}
   \lambda_\Gamma = \sum\nolimits_\nu \lambda_{\Gamma\nu}
   \label{lmd_Gama}
\end{equation}
where $\sum\nolimits_\nu$ indicates the summation of all modes at zone-center $\Gamma$.~$\lambda_{\Gamma\nu}$ is defined by
\begin{equation}
   \lambda_{\Gamma\nu} = \frac{\tilde{\omega}_{\Gamma\nu}^2-\omega_{\Gamma\nu}^2}{4\omega_{\Gamma\nu}^2}
   \label{lmd_Gama_mu}
\end{equation}
where the $\omega_{\Gamma\nu}$ and $\tilde{\omega}_{\Gamma\nu}$ are screened and unscreened phonon frequencies of mode $\nu$ at zone-center, respectively. The phonon frequencies were calculated with the single-cell and finite displacement method implemented in the Phonopy code \cite{43_togo2015first}. The displacement amplitude in the frozen-phonon calculations is 0.02~\AA. The convergence criterion of total energy is ${10}^{-8}$ eV.

The calculations of full Brillouin-zone EPC constants and \textit{T$_c$} of MgB$_2$, Li$_3$ZrB$_8$, and Ca$_3$YB$_8$ were performed based on density-functional perturbation theory (DFPT) \cite{27_baroni2001phonons} implemented in {Quantum ESPRESSO} code \cite{44_giannozzi2009quantum, 45_giannozzi2017advanced, 46_giannozzi2020quantum}. We used the ultra-soft pseudopotentials from the GBRV library \cite{47_garrity2014pseudopotentials}. After the convergence test, the plane-wave cut-off and the charge density cut-off were chosen to be 60 and 500 Ry, respectively. The reference DFPT calculation of dynamical matrix and EPC matrix elements in MgB$_2$ is based on the AlB$_2$-type primitive cell with the k mesh of 24×24×24 and the q mesh of 6×6×6. The DFPT calculations of Li$_3$ZrB$_8$ and Ca$_3$YB$_8$ were based on the 2×2×1 supercell, using the k of 12×12×24 and the q mesh of 3×3×6. The convergence threshold was $1\times{10}^{-12}$ Ry. The gaussian smearing of width was 0.01 Ry. 

The isotropic Eliashberg spectral function was obtained via the average over the Brillouin zone \cite{48_eliashberg1960interactions}
\begin{equation}
   \alpha^2F(\omega) = \frac{1}{2N(\epsilon_F)}\sum\nolimits_{\boldsymbol{q}\nu}\frac{\gamma_{\boldsymbol{q}\nu}}{\hbar\omega_{\boldsymbol{q}\nu}}\delta(\omega - \omega_{\boldsymbol{q}\nu})
   \label{lmd_Gama_mu}
\end{equation}
where $N(\epsilon_F)$ is the density of states at the Fermi level $\epsilon_F$; $\omega_{\boldsymbol{q}\nu}$ denotes the phonon frequency of mode $\nu$ with wave vector $\boldsymbol{q}$.
$\gamma_{\boldsymbol{q}\nu}$ is the phonon linewidth defined by 
${\gamma_{\boldsymbol{q}\nu} = \frac{2\pi\omega_{\boldsymbol{q}\nu}}{\Omega_{BZ}}\sum\nolimits_{ij}\int d^3 k|g_{\boldsymbol{k},\boldsymbol{q}\nu}^{ij}|^2 \delta(\epsilon_{\boldsymbol{q},i}-\epsilon_F)\delta(\epsilon_{\boldsymbol{k}+\boldsymbol{q},j}-\epsilon_F)}$,
where $g_{\boldsymbol{k},\boldsymbol{q}\nu}^{ij}$ is the EPC matrix element; $\epsilon_{\boldsymbol{q},i}$ and $\epsilon_{\boldsymbol{k}+\boldsymbol{q},j}$ are eigenvalues of Kohn-Sham orbitals at bands $i, j$ and wave vectors $\boldsymbol{q}$, $\boldsymbol{k}$. 

The full Brillouin-zone EPC constant $\lambda$ is determined through the integration of the Eliashberg spectral function
\begin{equation}
   \lambda = 2\int \frac{\alpha^2F(\omega)}{\omega}d\omega
   \label{lmd}
\end{equation}

The \textit{T$_c$} is obtained with the analytical McMillan equation \cite{49_mcmillan1968transition} modified by the Allen-Dynes \cite{50_allen1972neutron, 51_allen1975transition} 
\begin{equation}
   \textit{T$_c$} = \frac{\omega_{log}}{1.2}\exp[\frac{-1.04(1+\lambda)}{\lambda(1-0.62\mu^*)-\mu^*}]
   \label{Tc}
\end{equation}
where $\omega_{log}$ is the logarithmic average frequency $\omega_{log} = \exp[\frac{2}{\lambda}\int \frac{d\omega}{\omega}\alpha^2F(\omega)log\omega]$; $\mu^*$ is the effective screened Coulomb repulsion constant, set as 0.1.

\begin{acknowledgments}

Work at Guangdong University of Technology was supported by the Guangdong Natural Science Foundation of China (Grant No. 2017B030306003, and No.2019B1515120078). R. Wang was supported by the Guangdong Basic and Applied Basic Research Foundation (Grant No. 2021A1515110328 and 2022A1515012174). F. Zheng, Y. Fang and S. Wu were supported by the National Natural Science Foundation of China (11874307). C.Z. Wang, V. Antropov and F. Zhang were supported by the U.S. Department of Energy (DOE), Office of Science, Basic Energy Sciences, Materials Science and Engineering Division. Ames Laboratory is operated for the U.S. DOE by Iowa State University under Contract No. DE-AC02-07CH11358, including the grant of computer time at the National Energy Research Supercomputing Center (NERSC) in Berkeley. Y. Sun was supported by National Science Foundation Awards No. DMR-2132666. 
\end{acknowledgments}

\bibliography{ref}

\end{document}